# A Sunlight-pumped Two-dimensional Thermalized Photon Gas

Erik Busley, Leon Espert Miranda, Christian Kurtscheid, Frederik Wolf, Frank Vewinger, Julian Schmitt, and Martin Weitz

*Institut für Angewandte Physik, Universität Bonn, Wegelerstr. 8, 53115 Bonn, Germany*

The Liouville theorem states that the phase-space volume of an ensemble in a closed system remains constant. While gases of material particles can efficiently be cooled by sympathetic or laser cooling techniques, allowing for large phase-space compression upon suitable coupling to the environment, for light both the absence of an internal structure, as well as the non-conservation of photons upon contact to matter imposes fundamental limits for three-dimensional light-harvesting systems, such as fluorescence-based light concentrators. An advantageous physical situation can in principle be expected for dye-solution filled microcavities with a mirror spacing in the wavelength range, where low-dimensional photon gases with non-vanishing and tunable chemical potential have been experimentally realized.

Motivated by the goal to observe phase-space compression of sunlight by cooling the captured radiation to room temperature, in this work we theoretically show that in a lossless, harmonically confined system the phase-space volume scales as $(\Delta x \Delta p/T)^d$ = constant, where $\Delta x$ and $\Delta p$ denote the position and momentum spread and $d$ the dimensionality of the system ($d$=1 or 2). Experimentally, we realize a corresponding sunlight-pumped dye microcavity and demonstrate thermalization of scattered sunlight to a two-dimensional room temperature photon gas with non-vanishing chemical potential. Prospects of phase-space buildup of light by cooling, as can be feasible in systems with a two- or three-dimensional band gap, range from quantum state preparation in tailored potentials up to technical applications for more efficient collection of diffuse sunlight.

**I.) Introduction**

Compressing the phase space by harnessing coupling to the environment is of interest in a wide range of fields, ranging from applications of laser cooled atoms to solar light collection [1-5]. Along these lines, motivated most prominently by the collection of diffuse sunlight, techniques of non-imaging optics have been developed, aiming at guiding or concentration of light without the requirement to form an image of the source, and involving phase-space densities much below unity, i.e. in a purely classical domain [5]. Spatial light concentration has been observed in luminescent collectors based on dye-doped transparent plates capturing the emission of the optical absorbers falling in a certain emission cone by total internal reflection [4, 6]. In some experiments, upon increased reabsorption, the blue wing of the emitted spectrum approaches a blackbody-like spectrum, which illustrates the importance of thermodynamics to light concentration [7].

In blackbody radiation, which is the most common example among thermal radiators emitting into free space, the radiation density $S$ follows the Stefan-Boltzmann law $S \propto T^4$, correspondingly it rapidly diminishes with decreasing temperature $T$, as understood from a non-conservation of the photon number, and the chemical potential vanishes [8]. A lowering of the temperature in such a system will decrease, rather than enhance, the (optical) phase-space density. In contrast, optical quantum gases, which are subjected to light-confining structures on the wavelength scale to yield effectively one- or two-dimensional systems, have been demonstrated to provide an important prerequisite for cooling, namely independent tuning of particle number and temperature [9]. To date, this tunability has allowed for the realization of photon and polariton condensates, where at sufficiently low temperature and large densities particles condense into a macroscopically occupied ground state owing to quantum statistics [10-16].

In this work, we examine non-imaging optics for light concentration in a two-dimensional system. The reduced dimensionality realizes a low-frequency cutoff, providing a non-trivial ground state, and the mirror topography induces a trapping potential for cavity photons [16]. By repeated absorption re-emission cycles photons are sympathetically cooled to the temperature of the dye solution, which is at room temperature, resulting in a compression of the optical cloud in both position and momentum space. We theoretically show that in a lossless system, upon cooling light in a material-filled optical microcavity phase-space buildup of the collected radiation is well expected, and we give scaling laws for the expected compression. To experimentally investigate the possibility of solar light collection with a low-dimensional photon gas, we have performed a proof-of-principle experiment that is not optimized for high photon collection efficiency realizing a sunlight-pumped dye microcavity. Following up on earlier work using laser-based optical pumping [9], we here observe a thermalized two-dimensional photon gas with non-vanishing chemical potential employing the3urvaturt-based pumping arrangement. The spectral data of the cavity emission is in good agreement with thermodynamic expectations over the full emission bandwidth.

## II.) Experimental principle: Collecting photons in flatland

Our experiment uses an optical microcavity consisting of two highly reflecting curved mirrors spaced by a distance in the wavelength regime, filled with dye solution (Fig.1). The apparatus is an extension of a setup used in previous work [9, 10, 16]. The mirrors, due to their small spacing, introduce a low-frequency cutoff at $\hbar\omega_{\text{cutoff}} \simeq 2.1$ eV, see Fig. 2a, imprinting a spectrum of photon energies restricted to well above the thermal energy $k_B T \simeq 1/40$ eV, with $T \simeq 300$K being the temperature of the experimental apparatus. This prevents the thermal emission of (optical) photons by the dye molecules. Photons trapped in the microcavity thermalize to the (rovibrational) temperature of the dye (also at room temperature) by repeated absorption and re-emission processes. In the course of the thermalization the longitudinal modal quantum number of the cavity photons remains fixed, as for small cavity distance emission predominantly occurs into a single longitudinal mode. The remaining transverse degrees of freedom make the photon gas in the microcavity effectively two-dimensional. The optical dispersion relation becomes quadratic, as for a massive particle, and the4urvaturee of the cavity mirrors induces a trapping potential for photons, as understood from the transverse variation of the required wavelength to match the boundary conditions imposed by the cavity mirrors. Away from the optical axis light with shorter wavelength than in the trap center, corresponding to higher photon energies, fulfills the boundary condition. We thus expect that a cooling of the photon gas, besides reducing the transverse momentum spread, will lead to a shrinking of the cloud in position space, as illustrated in Fig.2c, similarly to the cooling of trapped atomic gases [2], or the transverse motion in particle accelerators [17].

Photons confined in the resonator in paraxial approximation are described by a dispersion of the form [9]

$$E(x,y,p_x,p_y) \cong m(c/n)^2 + \frac{p_x^2 + p_y^2}{2m} + \frac{1}{2}m\Omega^2(x^2+y^2), \tag{1}$$

resembling that of a (two-dimensional) harmonically confined system of massive particles, where $m = \hbar\omega_{\text{cutoff}}/(c/n)^2$ is an effective photon mass. Here $c$ denotes the speed of light in vacuum, $n$ the refractive index of the dye medium used in our experiment, and $\omega_{\text{cutoff}}$ $=2\pi c/\lambda_{\text{cutoff}}$, with $\lambda_{\text{cutoff}}$ as the cutoff wavelength. Further, $x$ and $y$ are spatial coordinates transversal to the cavity axis, $p_x$ and $p_y$ the transverse photon momentum components, and $\Omega$ the trapping frequency induced by the mirror curvature. At the here used low photon numbers, the phase-space density remains well below unity such that classical statistical mechanics prescribes the occupation densities in the thermalized case, yielding a Boltzmann distribution of photons above the low-frequency cutoff [9, 16].

**III.) Expectations for phase-space buildup upon cooling**

We are interested to derive the phase-space distribution of a harmonically confined two-dimensional photon gas at different temperatures. The thermal average of an observable $A$ in the classical regime is determined by Boltzmann statistics [18] and given by

$<A> = (1/Z) \int dxdy \int dp_x dp_y A(x,y,p_x,p_y) \exp[-E(x,y,p_x,p_y)/k_B T]$, with

$Z = \int dxdy \int dp_x dp_y \exp[-E(x,y,p_x,p_y)/k_B T]$ and $p_{x,y} = \hbar k_{x,y}$. For the rms cloud widths in position and momentum space, one then finds $\Delta r_i = \sqrt{<r_i^2>} = \sqrt{k_B T/m\Omega^2}$ and $\Delta p_i = \sqrt{<p_i^2>} = \sqrt{mk_B T}$ with $i=\{x,y\}$ respectively, which agrees with the equipartition theorem in terms of average potential and kinetic energies of $k_B T/2$ per degree of freedom. Upon cooling, the two-dimensional harmonically trapped photon cloud shrinks both in position and momentum space, as outlined in Fig.2c. We thus expect the phase-space volume to linearly decrease with temperature, as $\Delta x \Delta p_x = k_B T/\Omega$. Expressed in terms of the angle of a paraxial ray with respect to the optical axis, $\theta_i \cong p_i/p_z = p_i \lambda_{cutoff}/(2\pi\hbar)$, we obtain a universal thermodynamic scaling for the product of the area of the effective aperture $A$ and solid angle $W$ given by

$$\frac{\Delta x \Delta y \Delta p_x \Delta p_y}{T^2} \propto \frac{A \cdot W}{T^2} = const. \qquad (2)$$

Other than the thermodynamic scaling laws for light collection efficiencies previously derived for Boltzmann-like photon gases [6, 7], eq. (2) contains no material-dependent quantities due to full thermalization of the photonic degrees of freedom possible with the here presented method. We obtain the expected phase-space distribution in thermodynamic equilibrium, with $N$ as the total photon number,

$$\rho(x, y, p_x, p_y) = \frac{\Omega^2 N}{4\pi^2 k_B^2 T^2} \exp\left[-\left(\frac{m}{2}\Omega^2(x^2+y^2) + \frac{p_x^2 + p_y^2}{2m}\right)/k_B T\right], \quad (3)$$

which is a Gaussian distribution both in position and in momentum space. In radial coordinates, $r = \sqrt{x^2 + y^2}$, $p_r = \sqrt{p_x^2 + p_y^2}$, the expected position and momentum space distributions take the form $\rho(r) = 1/(2\pi\sigma_r^2)\exp(-r^2/2\sigma_r^2)$, and $\rho(p_r) = 1/(2\pi\sigma_p^2)\exp(-p_r^2/2\sigma_p^2)$, with $\sigma_r = \Delta r_i \equiv \sqrt{k_B T/m\Omega^2}$ and $\sigma_p = \Delta p_i \equiv \sqrt{mk_B T}$, respectively.

In the following, we consider the situation that incident (hot) radiation is cooled down to the apparatus temperature by radiative contact to the thermalization medium (i.e. dye molecules). This is a much more realistic scenario than a change of the temperature of the apparatus on a timescale of the photon lifetime in the microcavity (for which one would have to account for the temperature dependence of the ratio of photons and dye electronic excitations [19]). In the limit of small losses such that the system remains thermalized, lowering the temperature quadratically reduces the phase-space volume, and we accordingly expect a quadratic increase of the phase-space density. This is also directly seen from eq. (3), which yields a central phase space density $\rho(0,0,0,0) = \frac{\Omega^2 N}{4\pi^2 k_B^2 T^2}$. When cooling down 5800 K spectral temperature sunlight radiation in an idealized sunlight concentrator to room temperature (300K), we correspondingly estimate an achievable enhancement in phase space density of $(5800/300)^2 \cong 370$.

**IV.) Realizing a Sunlight-Pumped Two-Dimensional Photon Gas**

The used optical resonator is formed by two highly reflecting dielectric mirrors with $R=1$m spherical curvature spaced by $D_0 \simeq 1.3$μm. The microcavity is filled with rhodamine 6G dye dissolved in ethylene glycol, the quantum yield of this dye is $\simeq 95\%$ [20]. Prior to filling, the solution is filtered. In the liquid solution, sub-picosecond timescale collisions with solvent molecules, being much faster than the electronic transitions, rapidly alter the rovibrational state of the molecules. The used dye well fulfills the Boltzmann-like Kennard-Stepanov frequency scaling between the spectral profiles of absorption $\alpha(\omega)$ and emission $f(\omega)$, which for a small bandwidth can be written as $f(\omega)/\alpha(\omega) \propto \exp(-\hbar\omega/k_B T)$ [21]. By repeated absorption re-emission processes, this thermal character of the dye molecules is rapidly transferred onto the photon degrees of freedom.

As described above, the short spacing of the cavity mirrors, which induces a large transverse mode spacing, makes the photon gas two-dimensional, with the longitudinal mode number being fixed (to $q = 7$ here). Correspondingly, the contact to the dye heat bath is expected to drive the photon gas to a thermalized distribution above the low-frequency cutoff. Other than in a perfect photon box, thermalization in our experimental system is mainly limited by finite mirror reflectivity. Figure 2b gives the calculated reflectivity of the dielectric mirrors as a function of both wavelength and angle of incidence for unpolarized light, with blue (yellow) color code corresponding to near unity reflection (transmission) coefficient. For radiation propagating under small angles $\theta$ with respect to the optical axis, the reflectivity in the for cavity photons relevant wavelength range of 540-590 nm exceeds 99.995%, and center wavelength of the dielectric coating is at 550 nm. For larger angles of incidence, however, the mirror reflectivity is strongly reduced and shows an intricate wavelength- and angle-dependent reflectivity pattern that is typical for multilayer dielectric structures. Correspondingly, any isotropically emitted (spontaneous) radiation cannot fully be re-captured by the cavity, in stark contrast to the ideal photon box model system. In earlier work using a comparable experimental setup, the number of reabsorption events of a photon before being lost from the resonator has been determined to be 3.8(2.6) in this case near the condensate threshold [16].

In order to pump the dye microcavity with sunlight, incoming day light was directed over a mirror mounted on two orthogonally-oriented motorized rotation stages, as to allow for a compensation of the effective sun trajectory, and focused with a 50mm aperture achromatic lens onto an optical multimode fiber (200μm core diameter, 5m length). Before coupling to the fiber, the sunlight passes a color filter suppressing transmission of light of wavelengths above 550nm to minimize stray light in the detected wavelength range. Given that the dye absorption in the long-wavelength stopband of the filter is small, the magnitude of the collected signal is only weakly reduced. To allow for a compensation of the effect of the earth rotation, part of the collected sunlight is split off before the fiber and directed to a camera, allowing for an adjustment of the mirror tilt angle via a servo loop with a time constant of a few seconds, much slower than the 50ms camera exposure time. Without this servo loop, the image of the sun moved out of focus on a near-to-one-minute timescale. Behind the fiber, the emission is imaged onto the dye microcavity. At the position of the dye microcavity, the beam diameter of the incident, spectrally broadband sunlight irradiation is near 180μm and inclined at 42° with respect to the optical axis. At this angle of incidence, in the relevant 510-550 nm wavelength window where significant dye absorption is present roughly 40% of the incoming unpolarized light is transmitted into the cavity. The radiation enters the cavity mirror from outside of the cavity through a right-angle glass prism glued to the outside of the cavity mirror substrate, such that entry into the glass occurs at near normal incidence. The single-pass absorption efficiency of the thin dye film in the relevant wavelength range for a typical dye concentration of 1 mmol/l (which well allows for the thermalization of cavity photons) is of order 1%, which limits the collection efficiency accordingly.

We have experimentally studied thermalization of photons in the sunlight-pumped microcavity. Typical parameters are a cutoff wavelength of $\lambda_{cutoff} \simeq 587$nm, yielding an effective photon mass $m = \hbar\omega_{cutoff}/(c/n)^2 \simeq 7.7 \cdot 10^{-36}$ kg at a dye refractive index $n \simeq 1.44$, and a trapping frequency of $\Omega = (c/n)\sqrt{2/D_0 R} \simeq 2\pi \cdot 40$GHz. Figure 3 gives spectra of the cavity emission recorded by analyzing the transmission through one cavity mirror with a slitless optical spectrometer, for varying concentration of the dye solution. While for a low concentration, see e.g. the 0.03 mmol/l data, a near flat spectral distribution above the low-frequency cutoff is visible, while the experimental data recorded at a 1 mmol/l dye concentration well agrees with the expectations for a Boltzmann distributed energy spectrum above the cutoff (dashed line). The former data is understood to be due to fluorescence arising from single scattering events given that photons here leak out of the cavity before being reabsorbed; only for higher concentration the reabsorption rate $R_{abs} = \rho\sigma(\lambda)c/n$, with $\rho$ as the dye concentration and $\sigma(\lambda)$ as the cross section, becomes large enough that photons relax to a thermal distribution in the trapping potential [22, 23]. We however point out that we find the spectra recorded with the used slitless spectrometer at very low concentration of limited significance, as attributed to the for this data large spatial spread of the trapped fluorescing photon cloud. For the largest used dye concentration of 3 mmol/l, the agreement between theory and experiment is less accurate, as attributed to the reduced quantum efficiency of the dye at such high concentrations due to dimer formation, which is known to also modify the spectral properties [24]. A further effect of high concentrations is that photons perform scattering events at a rate becoming comparable to the trapping frequency, which, given the finite recapturing probability of the fluorescence photons, can lead to cavity loss prior to relaxation of the spatial degrees of freedom.

Next we have studied the microcavity emission in position and momentum space. Spatial distributions of the photon gas are obtained by direct imaging of the transmitted microcavity emission using a long working distance microscope of 0.42 numerical aperture onto an EMCCD camera (manufacturer Andor, model iXon Ultra 897). To observe the transverse momentum (i.e. far field) distribution of the photon gas, the cavity emission (following again the microscope objective) is directed to a secondary optical path that images the Fourier plane onto an ICCD camera (manufacturer Hamamatsu, model C9546-02-71 (intensifier), ORCA-Flash 4 (camera)). Due to the limited numerical aperture of the microscope, the outer 10% of the emitted light cone fall outside of our detection angle. Figures 4a,b show exemplary data recorded in the real space and Fourier planes, respectively. The recorded images exhibit a spherically symmetric profile, in accordance with theory expectations.

From the position and momentum space image data, we determine the four-dimensional phase-space distribution of the light in the microcavity together with the corresponding photon number for different dye concentration for a cavity cutoff of $\lambda_{cutoff} \simeq 587$nm, see the data in Table 1. The given data is normalized to the used dye concentration, in order to account for the different sunlight collection efficiencies, which due to the small single-pass absorption can well be assumed to scale linearly with concentration. Each data point corresponds an average of 3 (8 for the case of the lowest concentration data to increase the signal-to-noise ratio) images recorded with 50ms (spatial) or 100ms (momentum) integration time, see Figs. 4a,b for typical raw data, and Fig. 5 for the calibrated photon density profiles in position and momentum space. In Fig. 5, the data at the lowest concentration can be described by a quasi-thermal distribution at $T$=600K, a temperature considerably lower than the surface temperature of the sun. For the data with concentrations above 0.3 mmol/l, the phase-space distribution can be well described by a room temperature distribution (Table 1). Although the observed phase-space density does not increase as would be expected for a lossless system, we nevertheless observe a near constant phase-space density for concentrations up to 1 mmol/l despite the decrease in photon number with concentration.

We attribute the data at 1 mmol/l, see both the spectral, spatial and momentum space data (Figs. 3 and 5), as evidence for a thermalized two-dimensional photon gas of sunlight photons at room temperature. From the camera signals recorded at a concentration of 1 mmol/l we deduce a typical output power of $P_{\text{out}} = 26(10)\text{pW}$, corresponding to an average photon number in the microcavity of $N = 0.19(7)$. This is more than 5 orders of magnitude below the onset of Bose-Einstein condensation, for which a critical number of $N_c = \frac{\pi^2}{3}\left(\frac{k_B T}{\hbar\Omega}\right)^2$ ≅80,000 must be reached [16]. The given photon number corresponds to a chemical potential of $\mu = -12.4(5)k_B T$, as was calculated by numerically solving $N = \sum_i n_{\mu,T}(E_i)$, with the Bose-Einstein distribution factor $n_{T,\mu}(E_i) = g(E_i)/(\exp((E_i - \mu)/k_B T) - 1)$, where we sum over the system eigenstates. Here $g(E_i) = 2(E_i/\hbar\Omega + 1)$ denotes the energy degeneracy of modes in the microcavity. Because of $\mu \ll -k_B T$, the term -1 in the denominator of the Bose-Einstein distribution function can be neglected and one arrives at a Boltzmann distribution.

**V.) Conclusions**

To conclude, we have experimentally demonstrated a sunlight pumped two-dimensional photon gas with non-vanishing chemical potential using a dye-microcavity setting. Furthermore, we have given universal thermodynamic scaling laws for the scaling of the optical phase-space density with temperature of trapped low-dimensional photon gases.

For the future, it will be interesting to experimentally observe a phase-space buildup of light by cooling. A common challenge both in the fields of optical quantum gases as well as light concentrators is the minimization of photon loss. While the present microcavity experiment relies on a one-dimensional bandgap, future work could exploit systems with a three-dimensional band gap, e.g. material-filled photonic crystal cavities [25, 26], to minimize losses and overcome limitations of the Liouville theorem. Such three-dimensional geometries could also allow employing absorber materials with larger optical depth to enhance the light collection efficiency. A trapping potential to confine the photon cloud in a photonic crystal may be achieved by employing spatial gradients in the band gap, which is a viable approach for light concentration in propagating geometries [27]. As a further perspective, microcavities or photonic crystals filled with quantum dots of high quantum efficiency can, besides an improved photo-stability, allow for a more broadband absorption as compared to dye molecules [28], which can then well extend into the ultraviolet spectral regime.

**Acknowledgements:**

This work was funded by the DFG within the focused research center SFB/TR185 (277625399) and the Cluster of Excellence ML4Q (EXC 2004/1 – 390534769), the EU within the PhoQuS consortium (No. 820392), and the DLR with funds provided by the BMWi (50WM1859). J.S. acknowledges support by an ML4Q Independence Grant.

**Figure Captions:**

Fig. 1: In the used setup, two highly reflective mirrors form a cavity of mirror spacing $D_0 \cong 7\lambda/2$ on the order of the optical wavelength, which is filled with a liquid dye solution. Incoming sunlight is scattered from the dye molecules, populates the cavity and becomes concentrated. The intensity distribution of the light trapped inside the cavity exhibits a mean spatial width $\Delta x$, and the emission transmitted through the cavity mirrors (here: bottom) diverges with an angle spread $\Delta\theta$. In the experiment, the sunlight is guided by an optical fiber for more stable coupling to the cavity.

Fig. 2: (a) Absorption and emission spectra of rhodamine 6G dye, along with the wavelength $\lambda_{cutoff}$, which acts as a cavity low-frequency cutoff. The solid vertical lines indicate transverse cavity modes. (b) Mirror transmission as a function of wavelength and angle of incidence for unpolarized light and rays incident from the side of the dye medium, assuming the cavity is filled with ethylene glycol (the dye solvent) with refractive index $n=1.44$. The mirrors are highly reflective for dye emission wavelengths close to normal incidence (0°) but become transmissive under large emission angles, which occur for isotropic fluorescence. (c) Illustration of the expected compression of the photon cloud within the trapping potential in both spatial and angular spread as the photons are sympathetically cooled from their initial, hot distribution (brown) to a room temperature distribution (green) by radiative contact to the dye.

Fig. 3: Spectra of the cavity emission at a cutoff wavelength $\lambda_{cutoff} = 587$nm for different dye molarities. The occupations are normalized by the absorbance (proportional to molarity) for better comparison. For higher dye molarities, reabsorption events become more likely, and the emission is approaching a thermal distribution below the cutoff wavelength, the dashed line

gives a Bose-Einstein distribution of modes at 300K. Due to the weak absorbance at low molarities the noise level is significantly higher for those measurements.

Fig. 4: (a) Spatial and (b) Fourier plane images of the microcavity emission, providing raw data for the real and momentum space distributions of the photons in the resonator. The profile cuts shown at the bottom refer to the positions of the dashed white lines. The visible steps in (b) at around ±23° are caused by the finite aperture of the imaging system.

Fig. 5: Radial profiles of the photon density distribution in position and momentum space (the latter expressed in terms of angular spread) for three dye molarities along with fitted Gaussian curves.. The distributions have been normalized to the area at a molarity of 1mmol l$^{-1}$, Above a molarity of 0.3 mmol/l the measured width of $\sigma_r$ and $\sigma_p$ are close to the expectation for a room temperature distribution of $\sigma_r \approx 92\mu m$ and $\sigma_p \approx 1.8 \times 10^{-28}$ kg m s$^{-1}$ (corresponding to an angular spread of $\sigma_\theta \approx 12.7°$), respectively, see Table 1.

| Concentration C (mmol l$^{-1}$) | 0.010 | 0.030 | 0.100 | 0.300 | 1 | 3 |
|---|---|---|---|---|---|---|
| Reabsorption time at $\lambda_{cutoff} = 587$nm (ns) | 26.4 | 8.8 | 2.64 | 0.88 | 0.264 | 0.088 |
| Photon number $N$ | 0.0071(5) | 0.017(1) | 0.0456(3) | 0.096(3) | 0.195(7) | 0.17(3) |
| Normalized photon number $N/C$ (mmol$^{-1}$ l) | 0.71(5) | 0.56(3) | 0.456(3) | 0.32(1) | 0.195(7) | 0.057(1) |
| $\sigma_r$ (µm) | 110(3) | 105(1) | 99.5(2) | 93.0(1) | 85.0(2) | 77.1(2) |

| $\sigma_p$ ($10^{-28}$ kg m s$^{-1}$) | 2.63(13) | 2.51(5) | 2.329(4) | 2.13(3) | 1.918(6) | 1.709(3) |
| --- | --- | --- | --- | --- | --- | --- |
| Normalized central phase space density $\rho(0,0,0,0)/C$ ($10^{61}$ m$^{-4}$ kg$^{-2}$ s$^2$ mmol$^{-1}$ l) | 2.14(29) | 2.05(14) | 2.15(2) | 2.07(8) | 1.86(6) | 0.83(1) |

Table 1: Extracted spatial cloud width $\sigma_r$ and momentum spread $\sigma_p$ together with the average photon number in the microresonator for different dye concentration, from which the central phase space density $\rho(x=0, y=0, p_x=0, p_y=0)$ is determined using Gaussian fits. Using the normalized photon number and central phase space density, the differing collection efficiencies for the incoming sunlight can be accounted for. For each dye concentration also the corresponding reabsorption time is given. For comparison, the photon lifetime in the microcavity, assuming for the sake of simplicity the reflectivity values at the cutoff wavelength (587 nm), varies between 200 ps for on-axis radiation ($\theta \cong 0°$), and clearly below a ps for large angles of incidence θ at which the dielectric mirrors become transmissive (see also Fig. 2b). Further, the oscillation period (i.e. the inverse trapping frequency) of photons in the trapping potential is $T = 2\pi/\Omega \cong 12.5$ ps. Errors are standard deviations from repeated measurements.

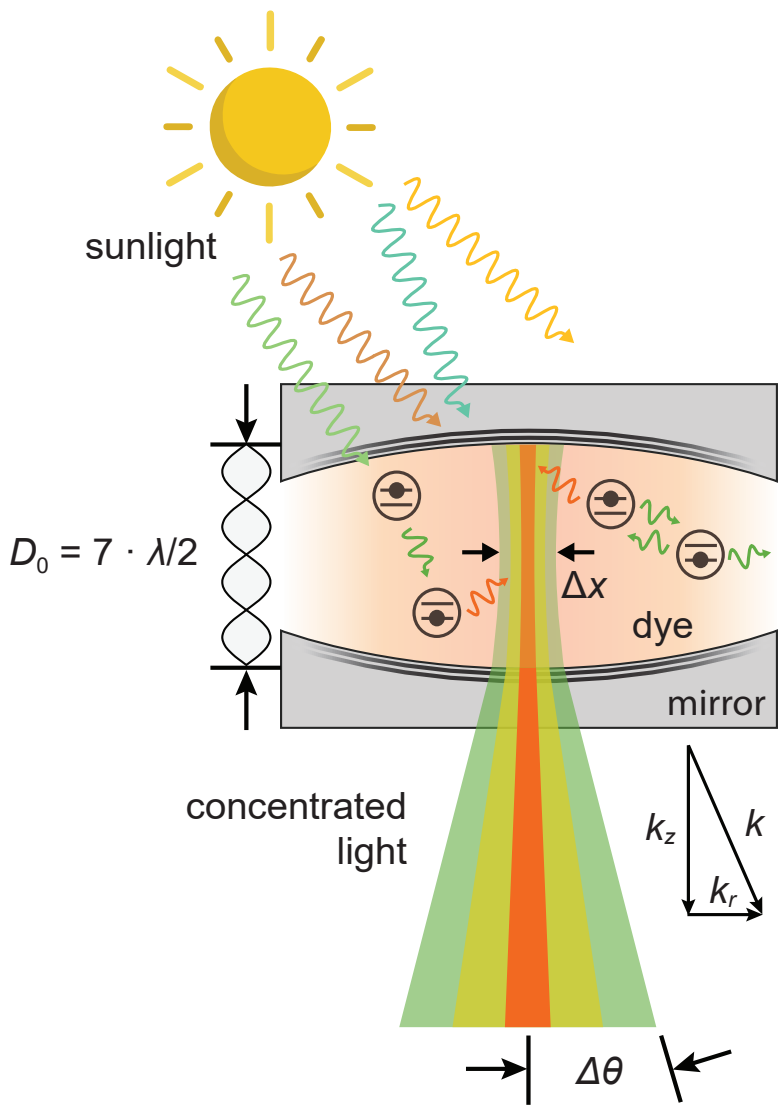

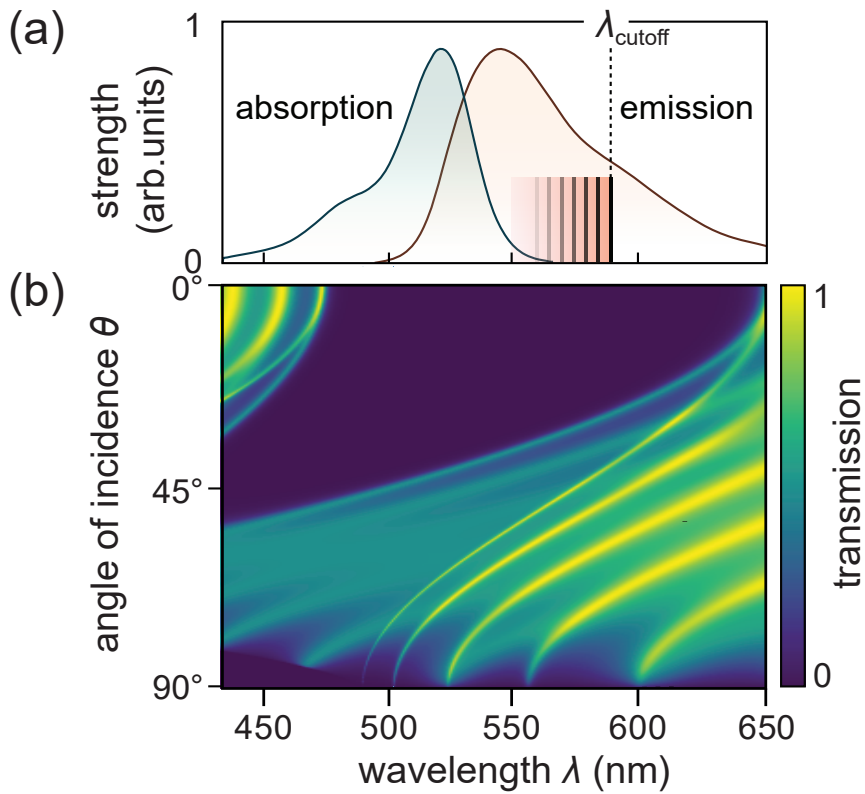

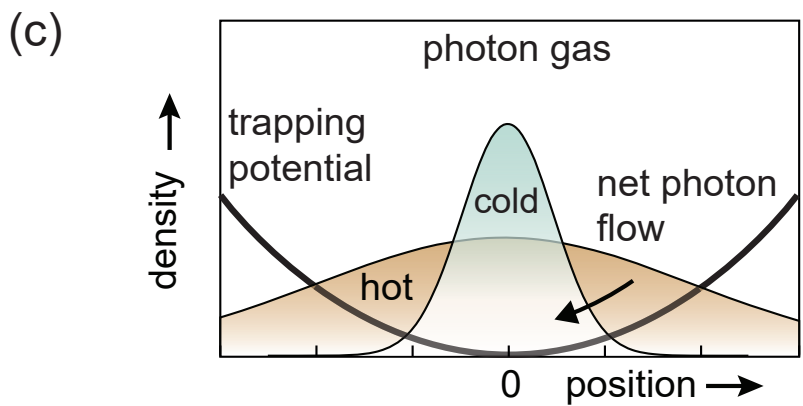

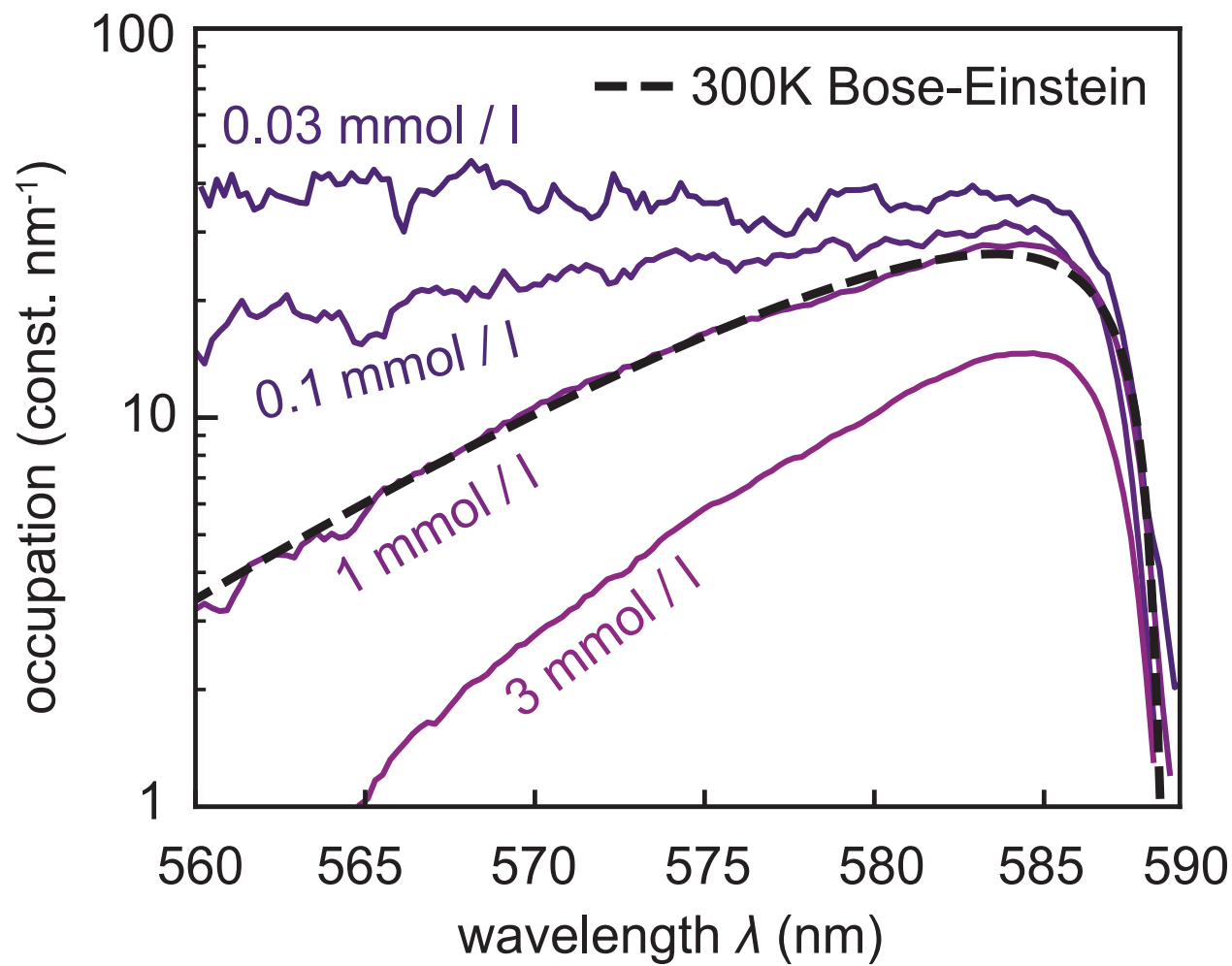

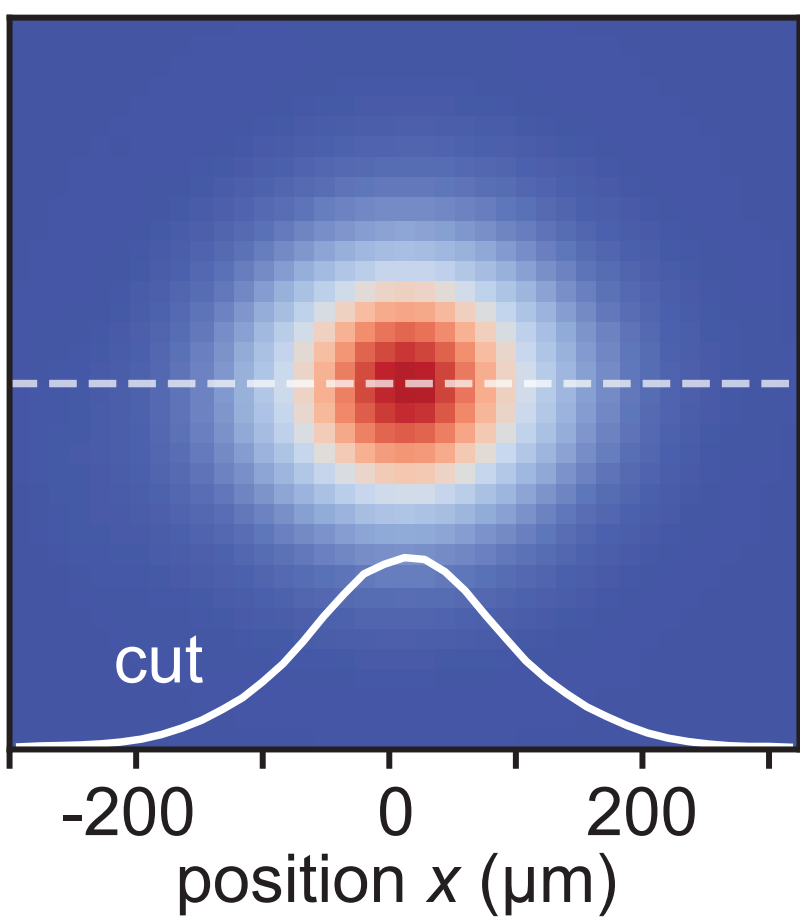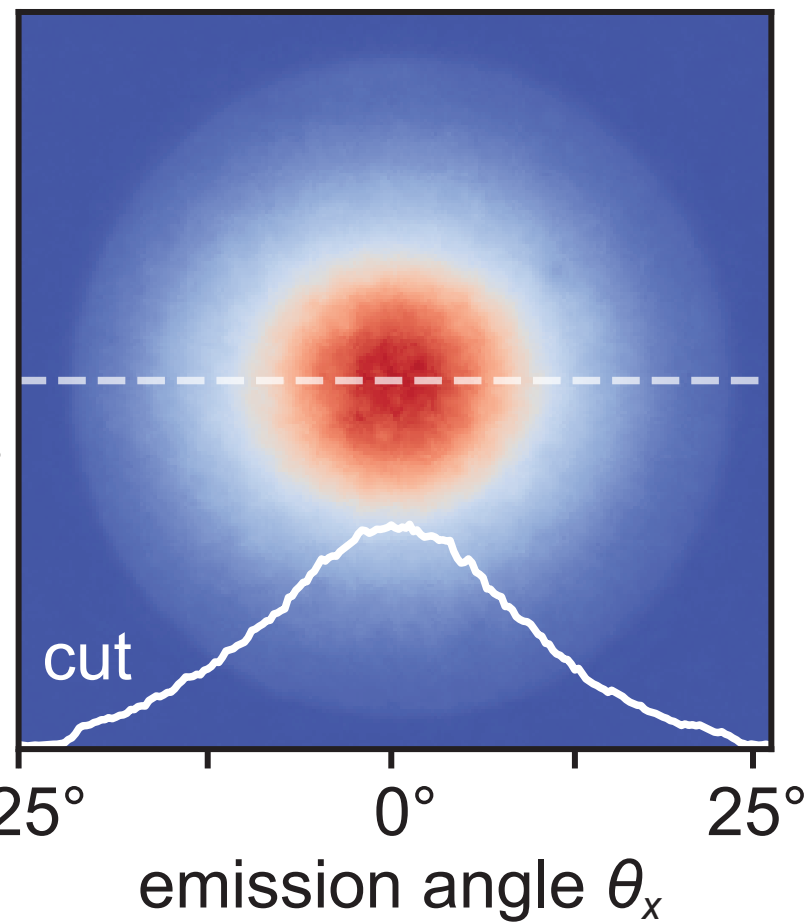

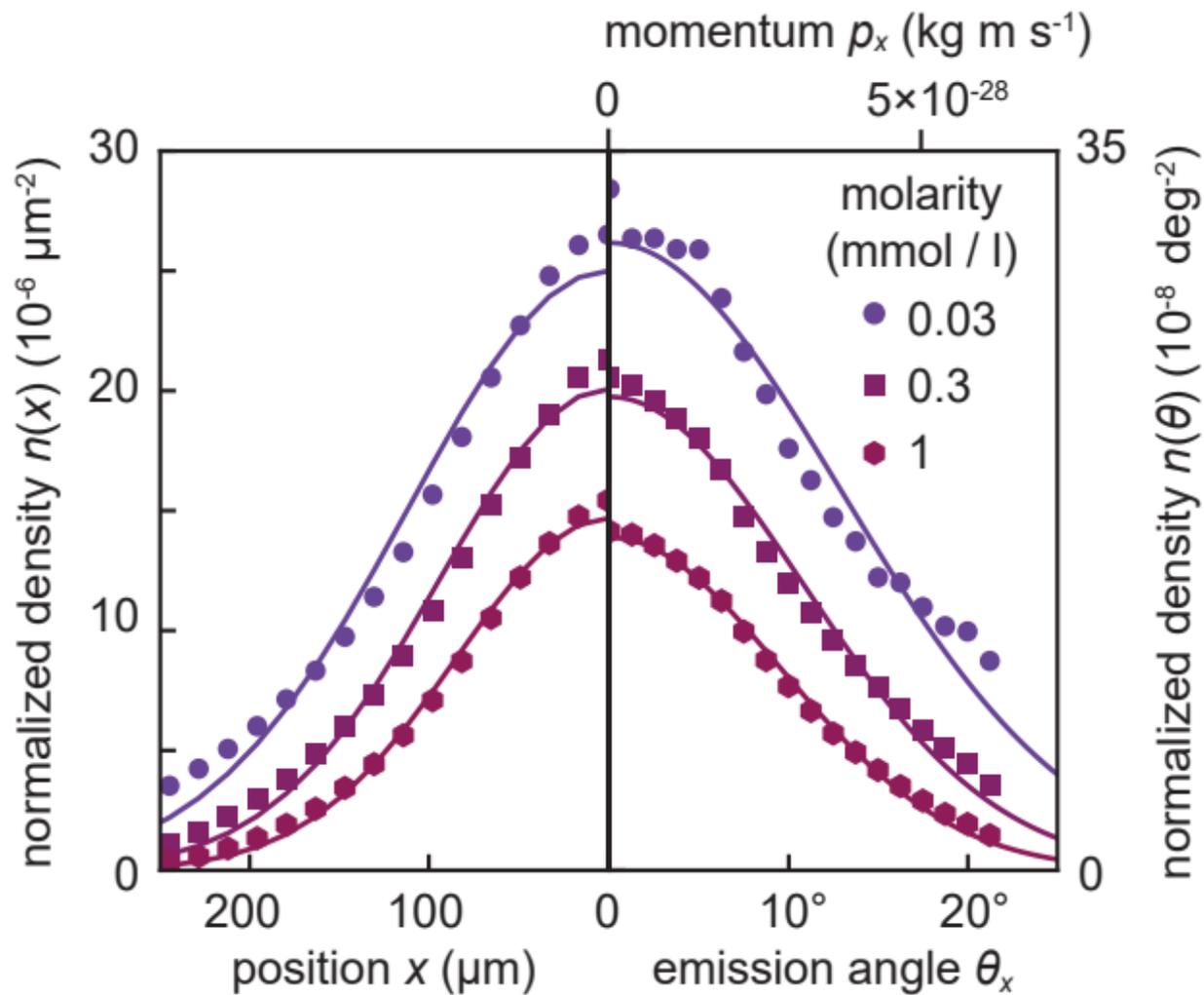